\documentstyle[aps,pra,eqsecnum,amsbsy,graphicx]{revtex}
\newcommand{\ve}[1]{{\boldsymbol{#1}}}
\begin{document}
\draft

\twocolumn

\wideabs{
\title{DYNAMICAL SUSCEPTIBILITIES IN STRONG COUPLING APPROACH:
GENERAL SCHEME AND FALIKOV-KIMBALL MODEL}
\author{A. M. Shvaika}
\address{Institute for Condensed Matter Physics,
National Academy of Sciences of Ukraine, \\
1~Svientsitskii Str., UA--79011 Lviv, Ukraine}

\maketitle

\begin{abstract}
A general scheme to calculate dynamical susceptibilities of strongly
correlated electron systems within the dynamical mean field theory is
developed. Approach is based on an expansion over electron hopping
around the atomic limit (within the diagrammatic technique for site
operators: projection and Hubbard ones) in infinite dimensions. As an
example, the Falicov-Kimball and simplified pseudospin-electron models
are considered for which an analytical expressions for dynamical
susceptibilities are obtained.

\end{abstract}

}

\section{Introduction}

The development of the Dynamical Mean Field Theory (DMFT), which is exact
in the $d=\infty$ limit, clarified some problems connected with the
simultaneous consideration of the electron hopping and strong local
correlations and stimulated a large progress in the understanding of the
strongly correlated electron systems~\cite{DMFTreview}. It was shown by
Metzner and Vollhardt~\cite{MetznerVoll,Metzner} that in the $d=\infty$
limit self-energies are single-site quantities (do not depend on wave
vector) which leads to a significant simplification. But such a local
self-energy only probes local properties in this limit and cannot detect
instabilities associated with a specific wave-vector~\cite{DMFTreview}
that requires the calculation of susceptibilities.

An analytical schemes developed for the descriptions of the strongly
correlated electrons can be divided into two types: (i) weak-coupling
theories, that are based on the expansion over local many-electron
interactions (Dyson approach) and are close to the standard Fermi-liquid
theory, and (ii) strong coupling theories, that start from the expansion
over electron hopping around the atomic limit (see Ref.~\cite{ShvaikaPRB}
and references therein). In the weak-coupling approach susceptibilities
are solutions of the Bethe-Salpeter equation and it is known that in the
$d\to\infty$ limit the corresponding irreducible four-vertices are also
local~\cite{DMFTreview,Zlatic}.

The aim of this article is to develop a general scheme to
calculate dynamical susceptibilities within a strong-coupling DMFT
approach for strongly correlated electron systems. A special case
of the binary alloy [Falicov-Kimball (FK)] type models, that can
be solved exactly in the limit of infinite
dimensions~\cite{BrandtMielsch}, will be also considered. Static
susceptibilities for the FK model have been already investigated
by Brandt and Mielsch~\cite{BrandtMielsch}, who found an
Ising-like phase transition to a chess-board charge-density-wave
phase at half filling, and Freericks~\cite{Freericks}, who showed
that system also displayed the incommensurate order and phase
separation at other fillings.

The preliminary short version of this article was published in the
Proceedings of M2S-HTSC-VI Conference~\cite{ShvaikaPC}. Here we
present the details of the general scheme.

\section{Two approaches in many-electron theory}

In general, the Hamiltonian of the electronic system with strong local
correlations can be written in the following form
\begin{equation}\label{Haml}
H=H_t+H_{\text{loc}},
\end{equation}
where
\begin{equation}
H_t = \sum_{ij\sigma} t_{ij} a^{\dag}_{i\sigma} a_{j\sigma}
\end{equation}
describes an intersite electron hopping and $H_{\text{loc}}$ is a sum
of the single site Hamiltonians
\begin{equation}
H_{\text{loc}} = \sum_i H_i
\end{equation}
which describe local electron correlations and/or interaction with
other local excitations (lattice vibrations, (pseudo)spins, etc.).

\subsection{Weak-coupling approach}

As a rule, the first term in (\ref{Haml}) is considered as an initial
(zero-order) Hamiltonian and the second one is considered as a
perturbation. Such approach is well known as the Dyson weak-coupling
approach, where the single electron Green's functions are determined by
the Dyson equation
\begin{equation}
G_{\sigma\ve k}(\omega_{\nu})\equiv
\raisebox{-6pt}{\includegraphics[width=5cm]{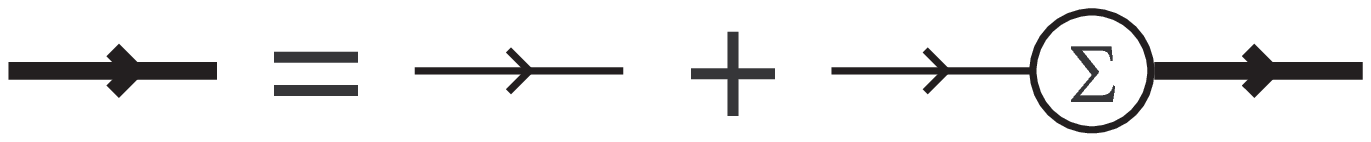}}
\end{equation}
or
\begin{eqnarray}\label{Dyson}
  G_{\sigma\ve k}(\omega_{\nu}) &=& {\cal G}_{\sigma\ve k}(\omega_{\nu}) +
  {\cal G}_{\sigma\ve k}(\omega_{\nu}) \Sigma_{\sigma\ve k}(\omega_{\nu})
  G_{\sigma\ve k}(\omega_{\nu})
  \\ \nonumber
  &=& \frac1{i\omega_{\nu} + \mu - t_{\ve k} - \Sigma_{\sigma\ve
  k}(\omega_{\nu})}.
\end{eqnarray}
Here
\begin{equation}
  {\cal G}_{\sigma\ve k}(\omega_{\nu})=\frac1{i\omega_{\nu} +
  \mu - t_{\ve k}}
\end{equation}
and the self-energy $\Sigma_{\sigma\ve k}(\omega_{\nu})$ describes all
scattering processes originated from the second term in (\ref{Haml}) and
cannot be divided into parts by cutting off one zero-order Green's
function ${\cal G}_{\sigma\ve k}(\omega_{\nu})$ line.

Also, it is well known that within such a weak-coupling approach
the electron susceptibilities (charge, spin, etc.) can be
presented in the form
\begin{equation}\label{D_suscept}
 \chi^{AB}_{\ve q}(\omega_m)=
 \raisebox{-11pt}{\includegraphics[width=4cm]{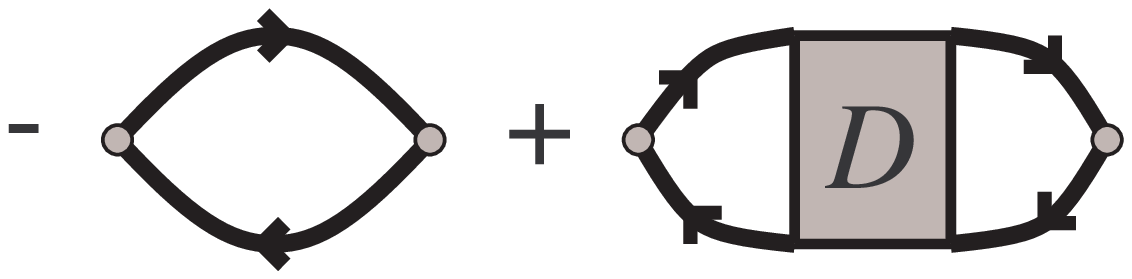}}\;,
\end{equation}
where the type of susceptibility is determined by the ending
parts. The full four vertex is a solution of the Bethe-Salpeter
equation
\begin{equation}\label{BS_eq}
 \raisebox{-10pt}{\includegraphics[width=4cm]{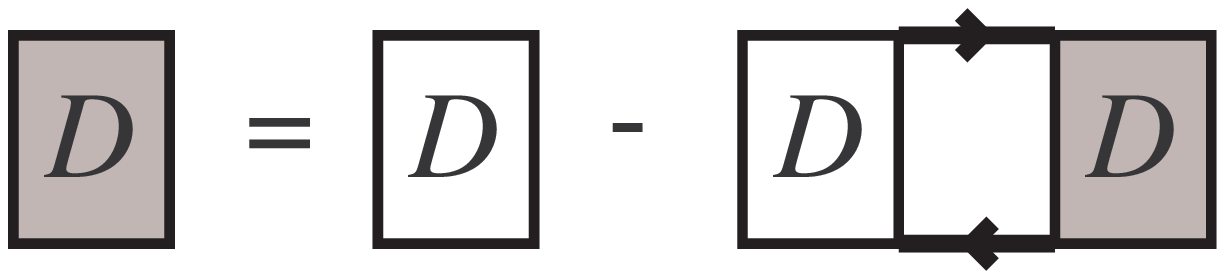}}
\end{equation}
with irreducible four vertex \framebox{\em D} that, in a same way as
self-energy, describes scattering processes originated from local
correlations and cannot be divided into parts by cutting off two
zero-order Green's function lines.

\subsection{Strong-coupling approach}

On the other hand, an alternative approach based on the expansion over
electron hopping around the strong-coupling limit can be
built~\cite{Larkin,Vaks}. In this case, the single-electron Green's
functions are determined by the Larkin equation~\cite{Larkin,Vaks}
\begin{equation}
G_{\sigma\ve k}(\omega_{\nu})
\equiv\raisebox{-6pt}{\includegraphics[width=5cm]{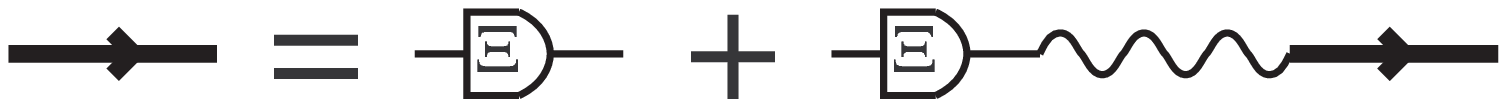}}
\end{equation}
or
\begin{eqnarray}\label{Larkin}
  G_{\sigma\ve k}(\omega_{\nu}) &=& \Xi_{\sigma\ve k}(\omega_{\nu}) +
  \Xi_{\sigma\ve k}(\omega_{\nu})\; t_{\ve k}\; G_{\sigma\ve k}(\omega_{\nu})
  \\ \nonumber
  &=& \frac1{\Xi^{-1}_{\sigma\ve k}(\omega_{\nu})-t_{\ve k}},
\end{eqnarray}
where $\Xi_{\sigma\ve k}(\omega_{\nu})$ is an irreducible according to
Larkin part which cannot be divided into parts by cutting off one hopping
(wavy) line.

Now, susceptibilities can be presented in the following form
\begin{equation}\label{L-suscept}
 \chi^{AB}_{\ve q}(i\omega_m)=
 \raisebox{-8pt}{\includegraphics[width=5cm]{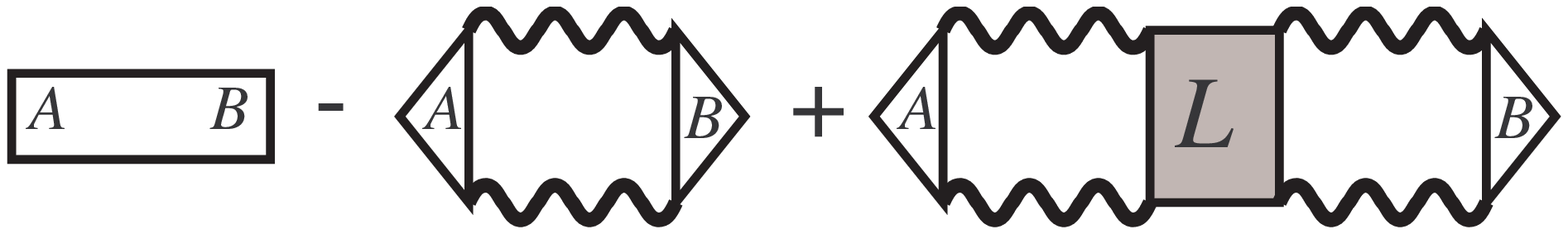}}\;,
\end{equation}
where
\begin{equation}\label{tchain}
\raisebox{-3pt}{\includegraphics[width=5cm]{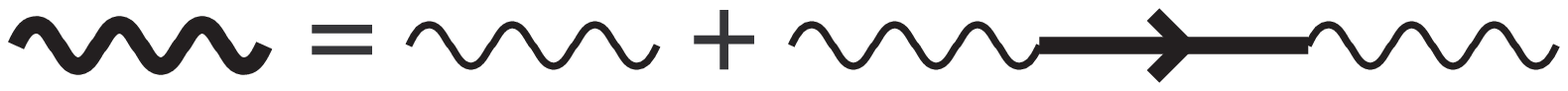}}
\end{equation}
is a sum of the chains of hopping lines and full four vertex is a
solution of the equation
\begin{equation}\label{L-BS}
 \raisebox{-10pt}{\includegraphics[width=4cm]{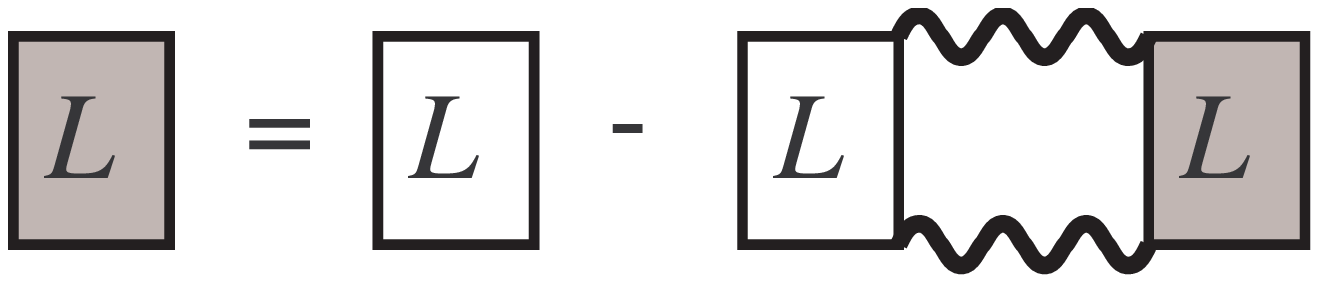}}\;.
\end{equation}
In Eqs. (\ref{L-suscept}) and (\ref{L-BS}) quantities
\frame{\phantom{OOO}}, $\Bigl\langle\hspace{-.28em}\Bigr|$ and
\framebox{\em L} are irreducible verticies which cannot be divided into
parts by cutting off two wavy (hopping) lines. It should be noted that, in
contrast to the weak-coupling approach, verticies in the strong-coupling
approach correspond to irreducible many-particle Green's functions.

\subsection{Connection between approaches}

From Eqs. (\ref{Dyson}) and (\ref{Larkin}) one can get a
connection between the self-energy and the irreducible vertex
function according to the Larkin equation:
\begin{equation}\label{cSigma}
 \Xi^{-1}_{\sigma\ve k}(i\omega_{\nu}) = i\omega_{\nu} + \mu -
 \Sigma_{\sigma\ve k}(i\omega_{\nu}).
\end{equation}

The connection between the four verticies in the weak and strong
coupling approaches is more complicated. For the full four
verticies one can get
\begin{equation}\label{c4full}
 \raisebox{-10pt}{\includegraphics[width=2.5cm]{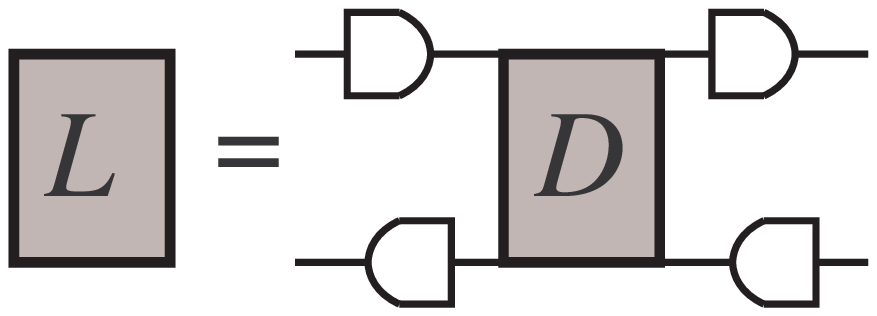}}\;,
\end{equation}
whereas the irreducible four verticies are connected by equation
\begin{equation}\label{c4irr}
 \raisebox{-10pt}{\includegraphics[width=2.5cm]{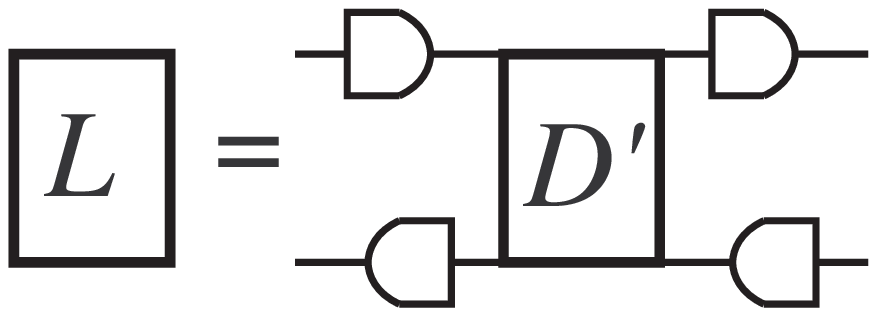}}\;,
\end{equation}
where \framebox{$D'$} is a solution of equation
\begin{equation}\label{c4prime}
 \raisebox{-10pt}{\includegraphics[width=4cm]{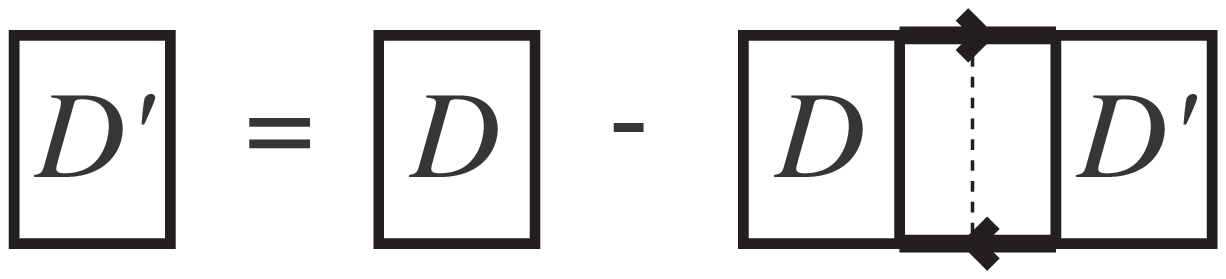}}\;.
\end{equation}
Here
\begin{equation}\label{lines}
 \raisebox{-10pt}{\includegraphics[width=6cm]{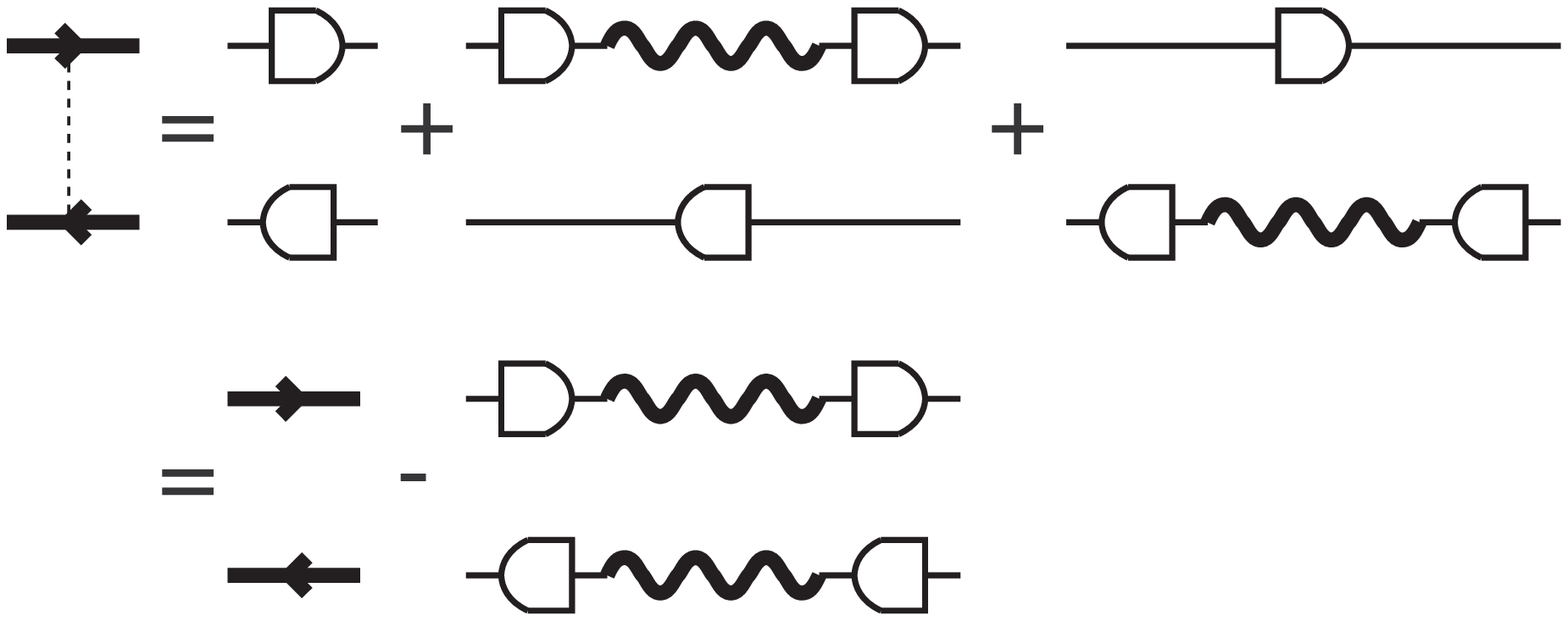}}
\end{equation}
In the case when the operators $\hat A$ and $\hat B$ are
constructed only by the electronic operators, then for three and
two verticies we find
\begin{equation}\label{c3irr}
 \raisebox{-12pt}{\includegraphics[width=3.5cm]{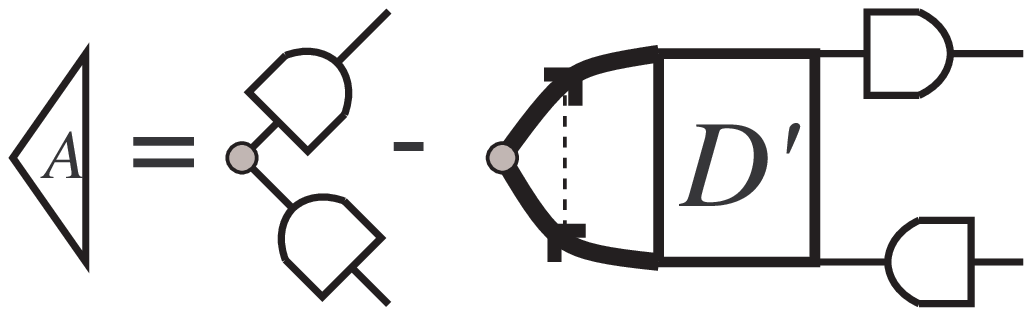}}
\end{equation}
and
\begin{equation}\label{c2irr}
 \raisebox{-10pt}{\includegraphics[width=4.5cm]{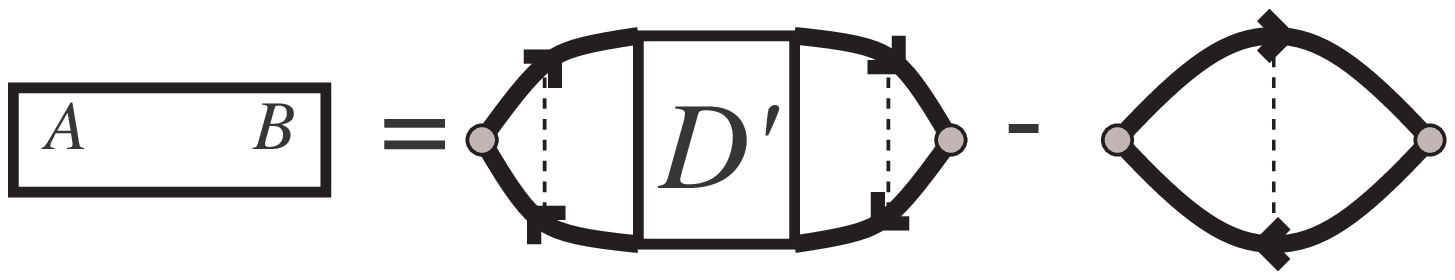}}
\end{equation}
which can be treated as some kind of sum rules.

\section{Dynamical mean field theory}

In the case of infinite dimensions $d\rightarrow\infty$ one should
scale the hopping integral according to
\begin{equation}\label{tinf}
t_{ij}\rightarrow \frac{t_{ij}}{\sqrt{d}}
\end{equation}
in order to obtain a finite density-of-states and it was shown by
Metzner~\cite{Metzner} that in this limit the irreducible part
becomes local
\begin{equation}
\Xi_{ij\sigma} (\tau-\tau')= \delta_{ij} \Xi_{\sigma} (\tau-\tau')
\end{equation}
or
\begin{equation}
\Xi_{\sigma\boldsymbol{k}}(\omega_{\nu}) = \Xi_{\sigma} (\omega_{\nu}).
\end{equation}
Such a site-diagonal function, as it was shown by Brandt and
Mielsch~\cite{BrandtMielsch}, can be calculated by mapping the
$d\rightarrow\infty$ lattice problem (\ref{Haml}) with intersite
hopping (\ref{tinf}) onto the atomic model with single-site
hopping
\begin{equation}\label{SIAM}
H_t \to \int_0^{\beta}d\tau \int_0^{\beta}d\tau' \sum_{\sigma}
\zeta_{\sigma}(\tau-\tau') a^{\dag}_{\sigma}(\tau)
a_{\sigma}(\tau').
\end{equation}
Here $\zeta_{\sigma} (\tau-\tau')$ is the auxiliary Kadanoff-Baym
field (dynamical mean field) which has to be self-consistently
determined from the condition that the same function
$\Xi_{\sigma}(\omega_{\nu})$ defines the Green's function for the
lattice and atomic model. The self-consistent set of equations for
$\Xi_{\sigma}(\omega_{\nu})$ and $\zeta_{\sigma}(\omega_{\nu})$
(e.g., see Ref.~\cite{DMFTreview} and references therein) is the
following:
\begin{eqnarray}
\frac{1}{N} \sum_{\boldsymbol{k}}
\frac{1}{\Xi_{\sigma}^{-1}(\omega_{\nu}) -t_{\boldsymbol{k}}} &=&
\frac{1}{\Xi_{\sigma}^{-1}(\omega_{\nu}) -
\zeta_{\sigma}(\omega_{\nu})}
\label{SIAMeq} \\
 &=&
G_{\sigma}^{(a)} (\omega_{\nu}, \{ \zeta_{\sigma} (\omega_{\nu})\}),
\nonumber
\end{eqnarray}
where
$G_{\sigma}^{(a)}(\omega_{\nu},\{\zeta_{\sigma}(\omega_{\nu})\})$
is the Green's function for the atomic model (\ref{SIAM}).

In the same way as it was done by Metzner~\cite{Metzner} for the
irreducible part $\Xi_{\sigma}(\omega_{\nu})$, one can prove that in the
$d\to\infty$ limit all irreducible verticies in the strong-coupling
approach (Eqs. (\ref{L-suscept}) and (\ref{L-BS})) are also single site
quantities. So, they can be also be calculated from the atomic model. On
the other hand, using Eqs. (\ref{cSigma})--(\ref{c2irr}) it is easy to
show that the self-energy and irreducible verticies in the weak-coupling
approach are also local quantities (in Ref.~\cite{Zlatic} that was proved
using different approach).

In order to calculate these single-site irreducible verticies first of all
one has to calculate an irreducible many particle Green's functions for
atomic model (\ref{SIAM}), i.e.
\begin{eqnarray}\label{4GF}
 \raisebox{-12pt}{\includegraphics[width=1cm]{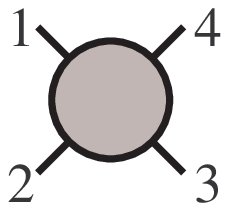}}&\equiv&
 \left\langle T a^{\dag}_1 a_2 \, a^{\dag}_3 a_4\right\rangle_{a}
 \\ \nonumber
  &-&\left\langle T a^{\dag}_1 a_2\right\rangle_{a}
     \left\langle T a^{\dag}_3 a_4\right\rangle_{a}
  + \left\langle T a^{\dag}_1 a_4\right\rangle_{a}
    \left\langle T a^{\dag}_3 a_2\right\rangle_{a},
\end{eqnarray}
\begin{equation}\label{3GF}
 \raisebox{-13pt}{\includegraphics[width=0.8cm]{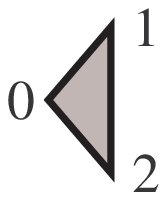}}\equiv
 \left\langle T \hat A_0 \, a^{\dag}_1 a_2 \right\rangle_{a}
 - \left\langle \hat A_0\right\rangle_{a}
   \left\langle T a^{\dag}_1 a_2\right\rangle_{a}
\end{equation}
and
\begin{equation}\label{2GF}
 \raisebox{-5pt}{\includegraphics[width=1.5cm]{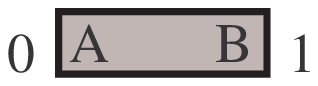}}\equiv
 \left\langle T \hat A_0 \, \hat B_1\right\rangle_{a}
 - \left\langle \hat A_0\right\rangle_{a}
   \left\langle \hat B_1\right\rangle_{a}.
\end{equation}

Then irreducible four vertex \framebox{$D$} for the weak-coupling approach
can be obtained from equation (\ref{BS_eq}), where now arrows indicate
single-electron Green's functions for the atomic model (\ref{SIAMeq}) and
full four vertex is determined from the many-particle Green's function by
\begin{equation}\label{4vsGF}
 \raisebox{-11pt}{\includegraphics[width=3cm]{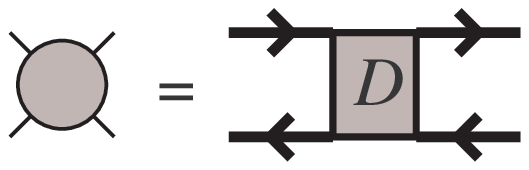}}\;.
\end{equation}

On the other hand, irreducible four vertex \framebox{$L$} for the
strong-coupling approach can be obtained from equation (\ref{L-BS}), where
the full four vertex is determined from the many-particle Green's function
by
\begin{equation}
 \raisebox{-8pt}{\includegraphics[width=4cm]{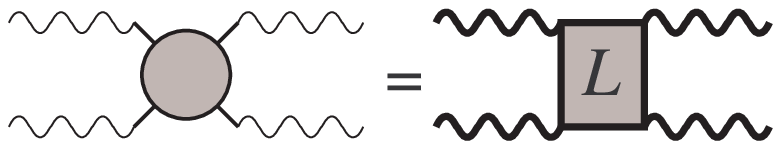}}
\end{equation}
or
\begin{equation}
 \raisebox{-9pt}{\includegraphics[width=4cm]{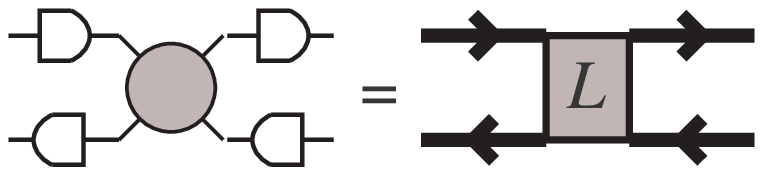}}\;.
\end{equation}
Now, the thin wavy lines are the dynamical mean field
$\zeta_{\sigma}(\omega_{\nu})$ and the thick ones, in contrast to
(\ref{tchain}), represent the sum of chains of wavy lines
\begin{equation}
\tilde \zeta_{\sigma}(\omega_{\nu})=
\frac{\zeta_{\sigma}(\omega_{\nu})}{1-\Xi_{\sigma}(\omega_{\nu})\zeta_{\sigma}(\omega_{\nu})}=
\zeta_{\sigma}(\omega_{\nu})\frac{G_{\sigma}^{(a)}(\omega_{\nu})}{\Xi_{\sigma}(\omega_{\nu})}.
\end{equation}

From the expression for the irreducible many particle Green's function
(\ref{3GF}), (\ref{2GF}) and(\ref{4GF}) for atomic model one can find
irreducible verticies $\Bigl\langle\hspace{-.28em}\Bigr|$ and
\frame{\phantom{OOO}} by equations
\begin{equation}
 \raisebox{-9pt}{\includegraphics[width=5cm]{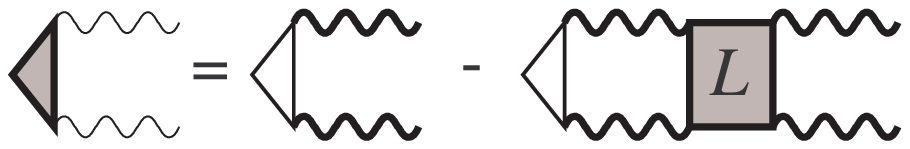}}
\end{equation}
and
\begin{equation}
 \raisebox{-9pt}{\includegraphics[width=6cm]{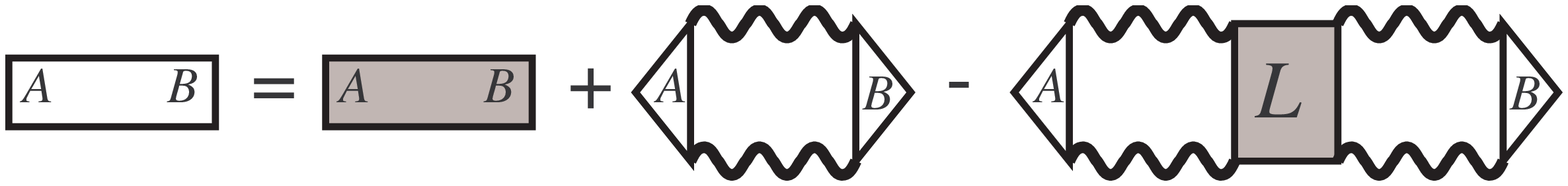}}\;,
\end{equation}
respectively, that complete the calculation of the irreducible
verticies for the dynamical susceptibilities.

Finally, for the lattice dynamical susceptibility in the weak
(\ref{D_suscept}) and strong (\ref{L-suscept}) coupling approaches we get
\begin{equation}
 \raisebox{-9pt}{\includegraphics[width=5.5cm]{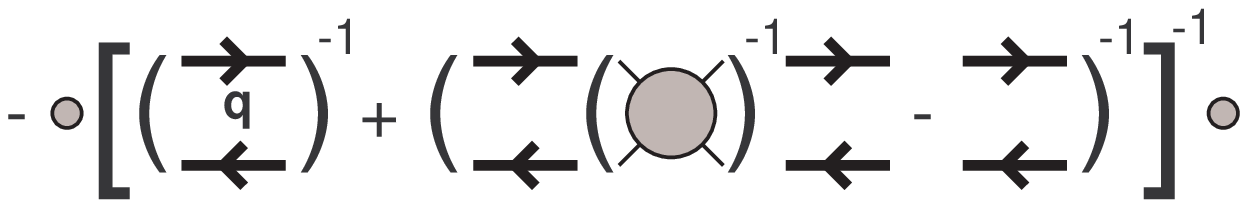}}\;,
\end{equation}
and
\begin{equation}\label{L-chi}
 \raisebox{-9pt}{\includegraphics[width=6.5cm]{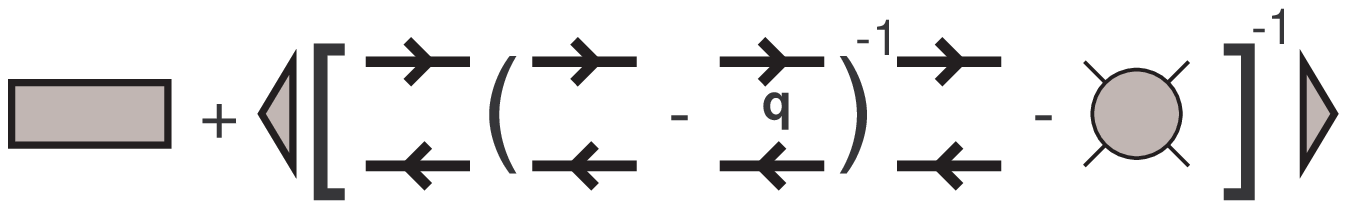}}\;,
\end{equation}
respectively. Here, $(\ldots)^{-1}$ denote an inverse kernels of the
corresponding integral equations,
\begin{equation}
 \raisebox{-10pt}{\includegraphics[width=0.6cm]{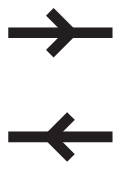}}=
 G_{\sigma}^{(a)}(\omega_{\nu})G_{\sigma}^{(a)}(\omega_{\nu+m}),
\end{equation}
and
\begin{equation}
 \raisebox{-10pt}{\includegraphics[width=0.6cm]{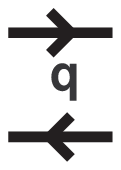}}=
 \frac1N \sum_{\ve k} G_{\sigma{\ve k}}(\omega_{\nu})G_{\sigma{\ve k+\ve q}}(\omega_{\nu+m}).
\end{equation}

\section{Binary alloy type models}

To test the possibilities of the above approach we consider a
binary alloy type model
\begin{equation}\label{H_FK}
H_i=\frac g2 \left(P^+_i - P^-_i\right) n_i - \frac h2 \left(P^+_i -
P^-_i\right), \label{binall}
\end{equation}
where $P^\pm_i = \frac 12 \pm S^z_i$ for the $U=0$ pseudospin-electron
(PE) model~\cite{PEM}, $P_i^+=c_i$, $P_i^-=1-c_i$ for a binary alloy, and
$a_{i\sigma}\to d_i$, $n_i=d^\dag_i d_i$, $P_i^+=f^{\dag}_i f_i$,
$P_i^-=1-f^{\dag}_i f_i$ for the Falicov-Kimball (FK)
model~\cite{Falicov}. The main difference between these models is in the
way that the averaging procedure is performed (a statistical one for the
PE and FK models and a configurational one for the binary alloy) and in
the thermodynamical regimes ($h={\rm const}$ for the PE model, $c={\rm
const}$ for the binary alloy and $n_d={\rm const}$ and/or $n_f={\rm
const}$ for the FK model).

The single-particle Green's function for the effective atomic model
(\ref{SIAM}) is a coherent sum of the Green's functions for subspaces
$S_i^z={\pm}\frac12$ and is equal to (see
Refs.~\cite{BrandtMielsch,JPS,ShvaikaPRB})
\begin{eqnarray}
  G_{\sigma}^{(a)}(\omega_{\nu}) &=&
  \frac{\left\langle P^+\right\rangle}
  {i\omega_{\nu} + \mu-\zeta_{\sigma}(\omega_{\nu})-\frac g2}
  \\ \nonumber
  &&+\frac{\left\langle P^-\right\rangle}
  {i\omega_{\nu} + \mu-\zeta_{\sigma}(\omega_{\nu})+\frac g2}
\end{eqnarray}
which allows us to find solutions of Eq.~(\ref{SIAMeq}) for the
irreducible part $\Xi_{\sigma}(\omega_{\nu})$ and dynamical mean field
$\zeta_{\sigma}(\omega_{\nu})$.

The Fourier transform of the two-electron Green's function $\left\langle T
a_1^{\dag}a_2a_3^{\dag}a_4\right\rangle$ is also a coherent sum of the
two-electron Green's functions for subspaces
\begin{eqnarray}
  &&G_{\sigma_1\sigma'_1}^{\sigma\sigma'}\left({\omega_{\nu}\atop\omega_{\nu+m}}
  {\omega_{\nu'}\atop\omega_{\nu'+m}}\right)=\sum_{\pm}\\
  &&\left\{\frac{\delta_{m0}\delta_{\sigma\sigma_1}\delta_{\sigma'\sigma'_1}
  \left\langle P^{\pm}\right\rangle}
  {\left[i\omega_{\nu} + \mu-\zeta_{\sigma}(\omega_{\nu})\mp\frac g2\right]
  \left[i\omega_{\nu'} + \mu-\zeta_{\sigma'}(\omega_{\nu'})\mp\frac
  g2\right]}\right.
  \nonumber\\
  &&\quad-\left.\frac{\delta_{\nu\nu'}\delta_{\sigma\sigma'}\delta_{\sigma_1\sigma'_1}
  \left\langle P^{\pm}\right\rangle}
  {\left[i\omega_{\nu} {+} \mu {-} \zeta_{\sigma}(\omega_{\nu}) {\mp} \frac g2\right]
  \left[i\omega_{\nu+m} {+} \mu {-} \zeta_{\sigma_1}(\omega_{\nu+m}) {\mp} \frac
  g2\right]}\right\}
  \nonumber
\end{eqnarray}
that gives for the irreducible two-electron Green's function
(\ref{4GF}) an expression which is a sum of the two separable functions
\begin{eqnarray}
  \tilde G_{\sigma_1\sigma'_1}^{\sigma\sigma'}\left({\omega_{\nu}\atop\omega_{\nu+m}}
  {\omega_{\nu'}\atop\omega_{\nu'+m}}\right)=&&
  \delta_{m0}\delta_{\sigma\sigma_1}\delta_{\sigma'\sigma'_1}
  \Lambda_{\sigma}(\omega_{\nu})\Lambda_{\sigma'}(\omega_{\nu'})
  \nonumber\\
  -\delta_{\nu\nu'}\delta_{\sigma\sigma'}&&\delta_{\sigma_1\sigma'_1}
  \Lambda_{\sigma}(\omega_{\nu})\Lambda_{\sigma_1}(\omega_{\nu+m}),
  \label{4GFirr}
\end{eqnarray}
where
\begin{equation}
  \Lambda_{\sigma}(\omega_{\nu})=\frac{g\sqrt{\langle P^+\rangle\langle
  P^-\rangle}} {\left[i\omega_{\nu} + \mu-\zeta_{\sigma}(\omega_{\nu})\right]^2-\frac{g^2}4}.
\end{equation}
Such a separable form allows us to calculate the inverse kernels
of the integral equations and to find all quantities in an
analytical form.

Within the weak-coupling approach, starting from the expression
(\ref{4GFirr}) with the use of Eq.~(\ref{4vsGF}), one can find solution of
the Bethe-Salpeter equation for the irreducible four vertex in an
analytical form:
\begin{eqnarray}\label{4vert-an}
  &&\Gamma_{\sigma_1\sigma'_1}^{\sigma\sigma'}\left({\omega_{\nu}\atop\omega_{\nu+m}}
  {\omega_{\nu'}\atop\omega_{\nu'+m}}\right)=
  \frac{\Lambda_{\sigma}(\omega_{\nu})}{G_{\sigma}^2(\omega_{\nu})+\Lambda_{\sigma}^2(\omega_{\nu})}
  \times\\
  &&\qquad\frac{\delta_{m0}\delta_{\sigma\sigma_1}\delta_{\sigma'\sigma'_1}}
  {1-\sum\limits_{\nu''\sigma''}\frac{\Lambda_{\sigma''}^2(\omega_{\nu''})}{G_{\sigma''}^2(\omega_{\nu''})+\Lambda_{\sigma''}^2(\omega_{\nu''})}}
  \cdot\frac{\Lambda_{\sigma'}(\omega_{\nu'})}{G_{\sigma'}^2(\omega_{\nu'})+\Lambda_{\sigma'}^2(\omega_{\nu'})}
  \nonumber\\
  &&-\frac{\delta_{\nu\nu'}\delta_{\sigma\sigma'}\delta_{\sigma_1\sigma'_1}}
  {G_{\sigma}(\omega_{\nu})G_{\sigma_1}(\omega_{\nu+m})}\times
  \nonumber\\
  &&\qquad\frac{\Lambda_{\sigma}(\omega_{\nu})\Lambda_{\sigma_1}(\omega_{\nu+m})}
  {G_{\sigma}(\omega_{\nu})G_{\sigma_1}(\omega_{\nu+m})+
  \Lambda_{\sigma}(\omega_{\nu})\Lambda_{\sigma_1}(\omega_{\nu+m})},
  \nonumber
\end{eqnarray}
which must be substituted into Eq.~(\ref{BS_eq}) for the full four vertex
for lattice and allows to calculate electron susceptibilities, e.g. for
charge susceptibility one can get
\begin{equation}\label{chi_nn}
\chi^{nn}_{\boldsymbol{q}}(\omega_m)=\delta_{m0}
\frac{\Delta^2_n}{T-\Theta(T,\boldsymbol{q})} +
K_{\boldsymbol{q}}^{nn}(\omega_m),
\end{equation}
\begin{equation}\label{K_nn}
  K_{\boldsymbol{q}}^{nn}(\omega_m)=\frac1{\beta}\sum_{\nu\sigma}
  \frac1{\chi_{\sigma\boldsymbol{q}}^{-1}(\omega_{\nu},\omega_m)-\Gamma_{\sigma}(\omega_{\nu},\omega_m)},
\end{equation}
where
\begin{eqnarray}\label{4vert}
\Gamma_{\sigma}(\omega_{\nu},\omega_m)&&=
\\ \nonumber
&&\frac{\Lambda_{\sigma}(\omega_{\nu})\Lambda_{\sigma}(\omega_{\nu+m})}
{\tilde\chi_{\sigma}(\omega_{\nu}\omega_m)
\left[\Lambda_{\sigma}(\omega_{\nu})\Lambda_{\sigma}(\omega_{\nu+m})-
\tilde\chi_{\sigma}(\omega_{\nu}\omega_m)\right]}
\end{eqnarray}
originates from the last term in (\ref{4vert-an}) and
\begin{eqnarray}
\Delta_n&&=
\\ \nonumber
&&\frac1\beta\sum_{\nu \sigma} \frac{\Lambda_{\sigma}
(\omega_{\nu})\tilde\chi_{\sigma}(\omega_{\nu}0)
\chi_{\sigma\boldsymbol{q}} (\omega_{\nu}0)}
{\tilde\chi^2_{\sigma}(\omega_{\nu}0)+\Lambda^2_{\sigma}(\omega_{\nu})
\left[\chi_{\sigma\boldsymbol{q}}(\omega_{\nu}0)-\tilde\chi_{\sigma}(\omega_{\nu}0)\right]},
\end{eqnarray}
\begin{eqnarray}\label{Theta}
\Theta(T,\boldsymbol{q})&&=
\\ \nonumber
&&\frac1\beta\sum_{\nu\sigma}
\frac{\Lambda^2_{\sigma}(\omega_{\nu})
\left[\chi_{\sigma\boldsymbol{q}}
(\omega_{\nu}0)-\tilde\chi_{\sigma}(\omega_{\nu}0)\right]}
{\tilde\chi^{2}_{\sigma}(\omega_{\nu}0)+\Lambda^2_{\sigma}(\omega_{\nu})
\left[\chi_{\sigma\boldsymbol{q}}(\omega_{\nu}0)-\tilde\chi_{\sigma}
(\omega_{\nu}0)\right]},
\end{eqnarray}
\begin{equation}
  \chi_{\sigma\boldsymbol{q}}(\omega_{\nu}\omega_m)=-\frac1N\sum_{\ve k}
  G_{\sigma\boldsymbol{k}}(\omega_{\nu})G_{\sigma\boldsymbol{k}+\boldsymbol{q}}(\omega_{\nu+m}),
\end{equation}
\begin{equation}
  \tilde\chi_{\sigma}(\omega_{\nu}\omega_m)=-
  G_{\sigma}^{(a)}(\omega_{\nu}) G_{\sigma}^{(a)}(\omega_{\nu+m}).
\end{equation}
But such a diagrammatic weak-coupling approach does not allow one
to calculate pseudospin and mixed susceptibilities.

On the other hand, the strong-coupling approach allows us to calculate all
susceptibilities. To do this, we calculate the irreducible many-particle
Green's functions
\begin{eqnarray}
&&\left\langle T n_{\sigma}(\tau) a_{\sigma'}^{\dag} (\tau')
a_{\sigma'} (\tau'') \right\rangle_{\text{irr}} \to
\delta_{m0}\sum_{\nu'}\Lambda_{\sigma}(\omega_{\nu'})\Lambda_{\sigma'}(\omega_{\nu})
\nonumber\\
&&-\delta_{\sigma\sigma'}\!\left[\Lambda_{\sigma}(\omega_{\nu})\Lambda_{\sigma}(\omega_{\nu+m}){+}
G_{\sigma}^{(a)}(\omega_{\nu})G_{\sigma}^{(a)}(\omega_{\nu+m})\right],
\label{naa}
\end{eqnarray}
\begin{eqnarray}
  \left\langle T S^z(\tau) a_{\sigma}^{\dag} (\tau')
  a_{\sigma} (\tau'') \right\rangle_{\text{irr}}&&\to
  \\ \nonumber
  && \delta_{m0}   \sqrt{\left\langle P^+ \right\rangle\left\langle P^-
  \right\rangle}\Lambda_{\sigma}(\omega_{\nu}),
\end{eqnarray}
\begin{equation}
\left\langle T S^z(\tau) S^z(\tau') \right\rangle_{\text{irr}} \to
\delta_{m0} \left\langle P^+ \right\rangle\left\langle P^-
\right\rangle,
\end{equation}
\begin{eqnarray}
&&\left\langle T n_{\sigma}(\tau) n_{\sigma'}(\tau')
\right\rangle_{\text{irr}}\! \to \delta_{m0}\!
\sum_{\nu}\Lambda_{\sigma}(\omega_{\nu})
\sum_{\nu'}\Lambda_{\sigma'}(\omega_{\nu'})
\\
&&-\delta_{\sigma\sigma'}\sum_{\nu}\left[\Lambda_{\sigma}(\omega_{\nu})\Lambda_{\sigma}(\omega_{\nu+m})+
G_{\sigma}^{(a)}(\omega_{\nu})G_{\sigma}^{(a)}(\omega_{\nu+m})\right]
\nonumber
\end{eqnarray}
\begin{eqnarray} \label{nn}
  \left\langle T S^z(\tau) n_{\sigma}(\tau')
  \right\rangle_{\text{irr}}&& \to
  \\ \nonumber
  && \delta_{m0} \sqrt{\left\langle P^+ \right\rangle\left\langle P^- \right\rangle}
  \sum_{\nu}\Lambda_{\sigma}(\omega_{\nu}),
\end{eqnarray}
After substituting (\ref{naa})--(\ref{nn}) into Eq.~(\ref{L-chi}) we get
for the charge susceptibility the same expression (\ref{chi_nn}), whereas
for the pseudospin and mixed susceptibilities
\begin{eqnarray}\label{chi_SS}
\chi_{\boldsymbol{q}}^{S^zS^z}(\omega_m)&=& \delta_{m0}
\frac{\Delta^2_{S^z}}{T-\Theta(T,\boldsymbol{q})}
\\ \nonumber
&=&\delta_{m0} \frac{\left\langle P^+ \right\rangle\left\langle
P^- \right\rangle}{T-\Theta(T,\boldsymbol{q})},
\end{eqnarray}
\begin{equation}\label{chi_nS}
\chi_{\boldsymbol{q}}^{nS^z}(\omega_m)=
\chi_{\boldsymbol{q}}^{S^zn}(\omega_m)= \delta_{m0}
\frac{\Delta_{S^z}\Delta_{n}}{T-\Theta(T,\boldsymbol{q})},
\end{equation}
where
\begin{equation}
\Delta_{S^z}=\sqrt{\left\langle P^+ \right\rangle\left\langle P^-
\right\rangle}.
\end{equation}

Expression (\ref{Theta}) for $\Theta(T,\boldsymbol{q})$ coincide with the
one obtained by Freericks~\cite{Freericks} from the equations for the
static susceptibilities ($\omega_m=0$) derived by Brandt and
Mielsch~\cite{BrandtMielsch}. Irreducible four vertex (\ref{4vert}) was
also derived by Freericks and Miller~\cite{FreericksMiller} within the
Baym-Kadanoff formalism and used to find the exact solution for the
nonresonant Raman scattering for the FK model~\cite{Raman1,Raman2}.

\section{Discussion}

For the binary alloy (Falikov-Kimball) model (\ref{H_FK})
expressions (\ref{chi_nn}), (\ref{chi_SS}) and (\ref{chi_nS})
define the so-called isothermal susceptibilities~\cite{Wilcox}.
Furthermore, the pseudospin (\ref{chi_SS}) and mixed
(\ref{chi_nS}) susceptibilities are only static (with factor
$\delta_{m0}$) because the pseudospin operator $S_i^z$ commutes
with the Hamiltonian and is an integral of motion. It should be
noted, that the binary alloy model (\ref{H_FK}) can be reduced to
the Ising type model with an effective multisite retarded
pseudospin interactions by taking trace of the statistical
operator with the Hamiltonian (\ref{H_FK}) over electron (fermion)
variables~\cite{ShvaikaPRB}. These explains the Ising type
expression obtained for the pseudospin susceptibility
(\ref{chi_SS}), but now the expression for the critical
temperature $\Theta(T,\boldsymbol{q})$ is more complicated.

The expression for the charge susceptibility (\ref{chi_nn})
contains two terms. The first one is static and can be written as
\begin{equation}
\chi^{nS^z}\left[\chi^{S^zS^z}\right]^{-1}\chi^{S^zn}.
\end{equation}
This describes the contribution from the pseudospin subsystem to
the charge susceptibility. It gives the main contribution to the
static susceptibilities. The second term
$K_{\boldsymbol{q}}^{nn}(\omega_m)$ in (\ref{chi_nn}) gives the
pure electron response and describes the so-called isolated (Kubo)
susceptibility~\cite{Wilcox}. In general, terms with the factor
$\delta_{m0}$ give the difference between the isothermal and
isolated susceptibilities (see Appendix in Ref.~\cite{Aksenov}).

\section*{Acknowledgement}

This work was partially supported by the Fundamental Researches State Fund
of the Ministry of Ukraine for Science and Education (Project
No~02.07/266).

\end{document}